\documentclass{ws-p8-50x6-00}
\newcommand {\pom} {I\!\!P}
\newcommand {\pomsub} {{\scriptsize \pom}}
\def\EPJC{{\em Eur. Phys. J.} C}
\begin{document}
%
%%%%%%%%%%%%%%%%%%%%%%%%%%%%%%%%% Title %%%%%%%%%%%%%%%%%%%%%%%%%%%%%%%%%%
%
\title{Study of $D^{*\pm}$ Meson Production \\ in 
                        Diffractive \lowercase {$ep$} Scattering at HERA}
%                                     Diffractive Scattering at HERA}
%
%%%%%%%%%%%%%%%%%%%%%%%%%%%%%%%%% Author %%%%%%%%%%%%%%%%%%%%%%%%%%%%%%%%%
%
\author{
I.A. Korzhavina (\lowercase {On behalf of the} 
                         \uppercase {ZEUS C}\lowercase{ollaboration})}
\address{        
Moscow State University, 119899 Moscow, Vorobievy Gory, Russia\\ 
          E-mail: irina@mail.desy.de}
%
%%%%%%%%%%%%%%%%%%%%%%%%%%%%%%%%%%%%%%%%%%%%%%%%%%%%%%%%%%%%%%
% You may repeat \author \address as often as necessary      %
%%%%%%%%%%%%%%%%%%%%%%%%%%%%%%%%%%%%%%%%%%%%%%%%%%%%%%%%%%%%%%
%
%%%%%%%%%%%%%%%%%%%%%%%%%%%%% Abstract %%%%%%%%%%%%%%%%%%%%%%%%%%%%%%%%%
%
\maketitle
%\centerline{\today}}
%
\abstracts{
       $D^{*\pm}$ production in the diffractive interactions  was measured  
with the ZEUS detector at HERA using integrated luminosities 
of 44 pb$^{-1}$ in deep inelastic scattering 
and 38 pb$^{-1}$ in photoproduction  processes.
%of 38 pb$^{-1}$ in photoproduction %(for the first time) 
%and 44 pb$^{-1}$ in deep inelastic scattering processes.
 Diffractive interactions were identified by a large gap
in the rapidity distribution of final states. 
$D^{*\pm}$ mesons were reconstructed
from the decay 
%channel
 $D^{*+} \to D^0\pi_s^+$ with \mbox{$D^0 \to K^-\pi^+$} (+c.c.). 
Integrated and differential cross sections were compared to 
theoretical expectations.}
%
%%%%%%%%%%%%%%%%%%%%%%%%%%%%% Introduction %%%%%%%%%%%%%%%%%%%%%%%%%%%%%%%%%
%

	$D^{*\pm}$ production was measured in the diffractive 
dissociation of the photon into a hadronic system $X$\cite{D*DIS00,D*PhP00}.
Photoproduction (PhP) reactions with photon virtualities $Q^2 < 1$ GeV$^2$ and 
 deep inelastic scattering (DIS) at $Q^2 > 4$ GeV$^2$ were studied separately.
   Kinematic regions for different cross sections were defined by 
$W$, $y$, $p_T(D^*)$, $\eta(D^*)$, $x_{\pomsub}$ and $\beta$, %variables. 
where $W$ is the photon-proton center-of-mass energy, 
$y$ is the fraction of the positron energy, carried  by the exchanged photon 
 in the proton rest frame, 
% transferred to the proton in its rest frame.
%transferred from positron to 
%the positron's momentum, taken by the photon.
$p_T(D^*)$ and $\eta(D^*)$ are the transverse momentum and pseudorapidity of
$D^{*\pm}$, respectively,
$x_{\pomsub}$ is the fraction of the proton's momentum carried by
the Pomeron ($\pom$) and $\beta$ is the  fraction of the Pomeron momentum 
participating in a hard subprocess.
%
%%%%%%%%%%%%%%%%%%%%%%%%%%%%% Signals %%%%%%%%%%%%%%%%%%%%%%%%%%%%%%%%%
%

Diffractive events were identified by a large rapidity gap (LRG) 
between the scattered proton and the hadronic system, $X$.
The LRG was tagged by a cut on $\eta_{max}$, the pseudorapidity of 
the most forward deposit of energy greater than 400 MeV in the detector.
%the most forward going particle of the event.
%
%The LRG events were separated from the non-diffractive ones 
% by a cut on $\eta_{max}$ at some value.
In PhP, the fraction of non-diffractive events, left after requiring
$\eta_{max} \leq 1.75$,  was estimated from a non-diffractive MC
%non-diffractive-MC-to-data distribution ratios
 and  subtracted.
	In DIS, the requirement $\eta_{max} \leq 2$ was used, 
where the non-diffractive admixture is negligible. 
The fraction of events with a dissociated proton, which was not rejected by the
cut on $\eta_{max}$, was estimated\cite{p.diss} to be 0.31$\pm$0.15 
 and  subtracted from all the measured cross sections.

	The integrated cross section for diffractive PhP of $D^{*\pm}$ in the 
kinematic region defined by $Q^2<1$ GeV$^2$, $130<W<280$ GeV,   
$p_T(D^*)> 2 $ GeV, $\vert\eta(D^*)\vert< 1.5$ and 
$0.001 < x_{\pomsub} < 0.018$   was found to be 
\centerline{$ \sigma(ep\to D^* Xp)_{PhP}^{diff} = 
0.74\pm 0.21(stat.)^{+0.27}_{-0.18}(syst.) \pm 0.16 (p. diss.)~nb.$ }
This value, while only a fraction of the total diffractive $D^{*\pm}$ 
contribution, corresponds to $\sim$ 4 $\%$ of the inclusive
 PhP\cite{incD*PhP} of $D^{*\pm}$ in the same kinematic range.  

	The integrated cross section for diffractive $D^{*\pm}$ production 
in DIS in the kinematic range defined by 
$4<Q^2<400$ GeV$^2$, $0.02<y<0.7$, 
$p_T(D^*)> 1.5 $ GeV, $\vert\eta(D^*)\vert< 1.5$,
$x_{\pomsub} < 0.016$ and $\beta < 0.8$
was measured to be 
\centerline{$ \sigma(ep\to D^* Xp)_{DIS}^{diff} = 0.281
\pm 0.041(stat.)^{+0.079}_{-0.073}(syst.)~nb. $ }
The fraction of
diffractively produced $D^{*\pm}$ was determined to be 
$\mathcal{R_D}=6.1\pm 0.9(stat.)^{+1.5}_{-1.4}(syst.) \%$,
consistent with the corresponding fraction for 
inclusive diffraction\cite{p.diss}.
The ratio is found to be independent of $Q^2$ and $W$. 	

The measured fractions
of diffractively produced $D^{*\pm}$ mesons indicate
that charm production is not suppressed in diffractive processes.

Both cross sections were compared with calculations, based on 
the resolved-Pomeron model\cite{resIP}, 
using the BGF mechanism for charm production %,
and  gluon-dominated 
 parameterizations of the Pomeron structure: the LO H1 FIT2\cite{H1FIT2LO} for PhP
  and the ACTW\cite{ACTW} for DIS. 
The PhP cross section, calculated with the RAPGAP MC\cite{RAPGAP}, is 1.42 nb, 
while  the calculated DIS cross section  
approximately coincides with the measured one. The calculations of the DIS
cross section using the quark-dominated parameterizations of the Pomeron 
structure do not reproduce the measurement.
%
%%%%%%%%%%%%%%%%%%%%%%%%%%%%% Differ Xsec %%%%%%%%%%%%%%%%%%%%%%%%%%%%%%%%%
%

	The shapes of the differential cross sections for $D^{*\pm}$ 
diffractive production (Figs~1~and~2) 
 are in qualitative agreement with the 
% resolved  Pomeron % resolved-Pomeron
 predictions based on the gluon-dominated Pomeron. 
%partonic distribution parametrizations within experimental errors.
The  DIS data also agree with the calculations in normalization. 
% DIS data shapes are
%in qualitative agreement with the two gluon model\cite{2gRIDI} as well 
%(Fig.~2).The data are also compared to the distributions of
% RIDI 
%% MC diffractive events, simulated in the frame of 
%using two gluon model
%%%Fig.~2 shows qualitative agreement in shapes between data and calculations(RIDI
%%%MC) based
%%%on the two gluon model\cite{2gRIDI} with real gluon corrections ($q\bar q g$).
%%%The leading order ($q\bar q$) calculations  based on the two
%%%gluon model fail to describe the $\beta$ distribution.
%%GLK 
Figure 2 shows that neither the $q\bar q$ nor the $q\bar q g$ component
 of the two-gluon model~\cite{2gRIDI} is able to describe the data
 adequately. However, an appropriate combination of the two
 components may well do so.
	The present statistics and systematics of the measurements do not 
discriminate between the models. 

%	I am grateful to the Organizing Committee of the Workshop DIS2001 and 
%Russian Academy of Science for financial support.
% of my participation in the Workshop.
%
%%%%%%%%%%%%%%%%%%%%%%%%%%%%%%% Bibliography %%%%%%%%%%%%%%%%%%%%%%%%%%%%%%%%%%%
%

%

%
\newpage

\begin{figure}[!h]
\parbox[t]{0.5\hsize}{\vglue0pt\resizebox*{\hsize}{!}
  {\includegraphics{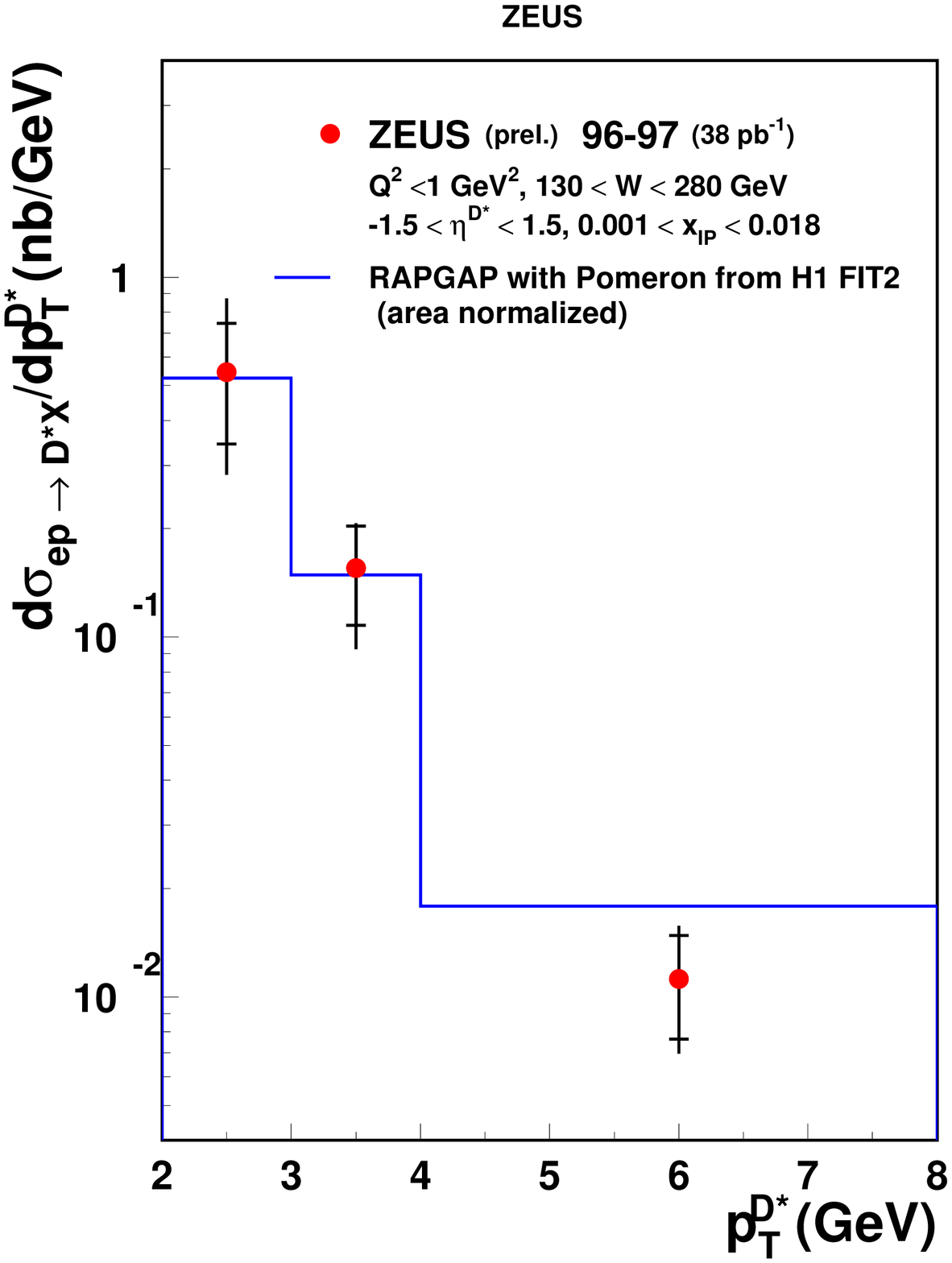}}
                     }%
\llap{\parbox[t]{27mm}{\vglue53mm\sffamily(a)}}
                                               \nolinebreak
\parbox[t]{0.5\hsize}{\vglue0pt\resizebox*{\hsize}{!}
  {\includegraphics{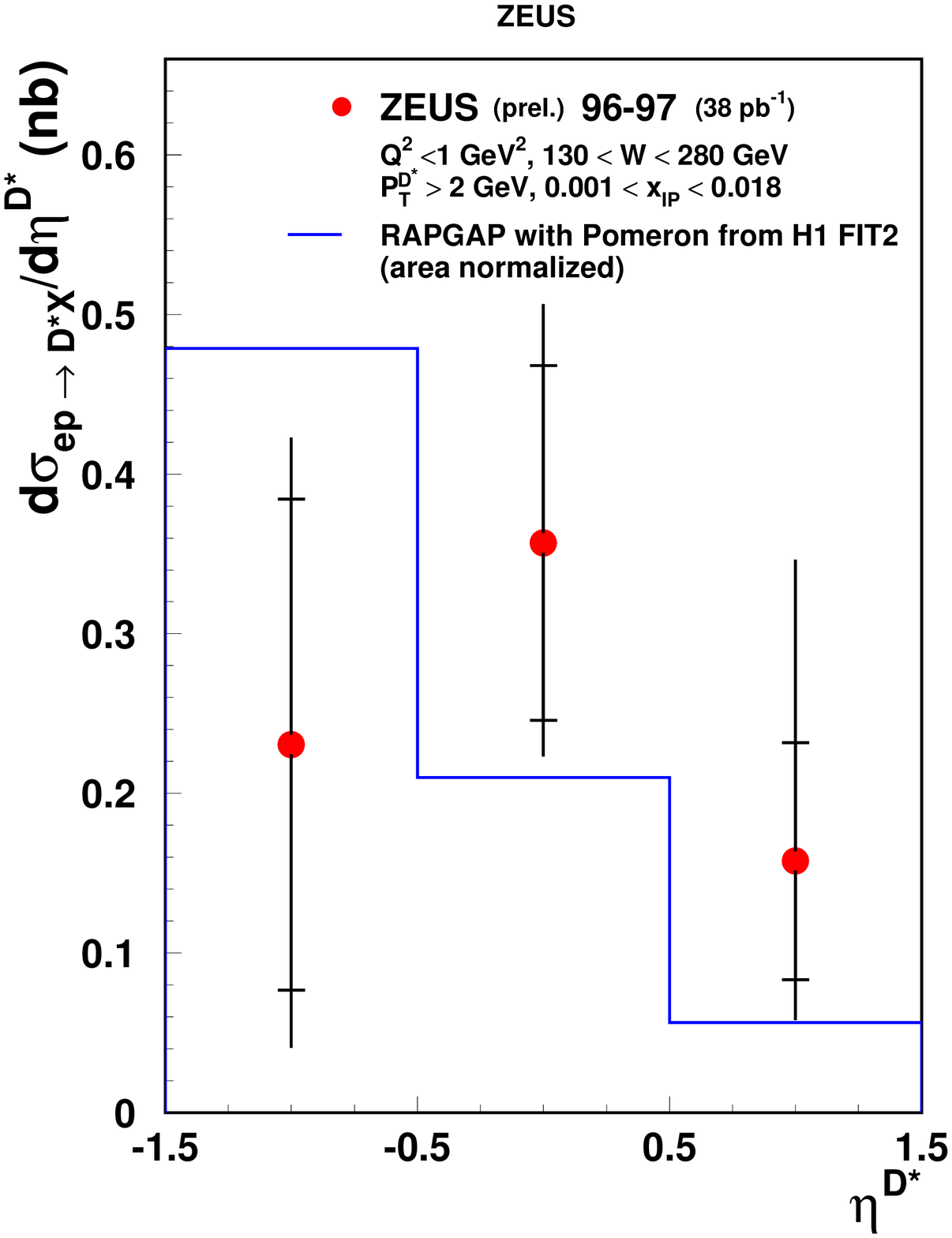}}
                     }%
\llap{\parbox[t]{27mm}{\vglue53mm\sffamily(b)}} \par
\parbox[t]{0.5\hsize}{\vglue0pt\resizebox*{\hsize}{!}
  {\includegraphics{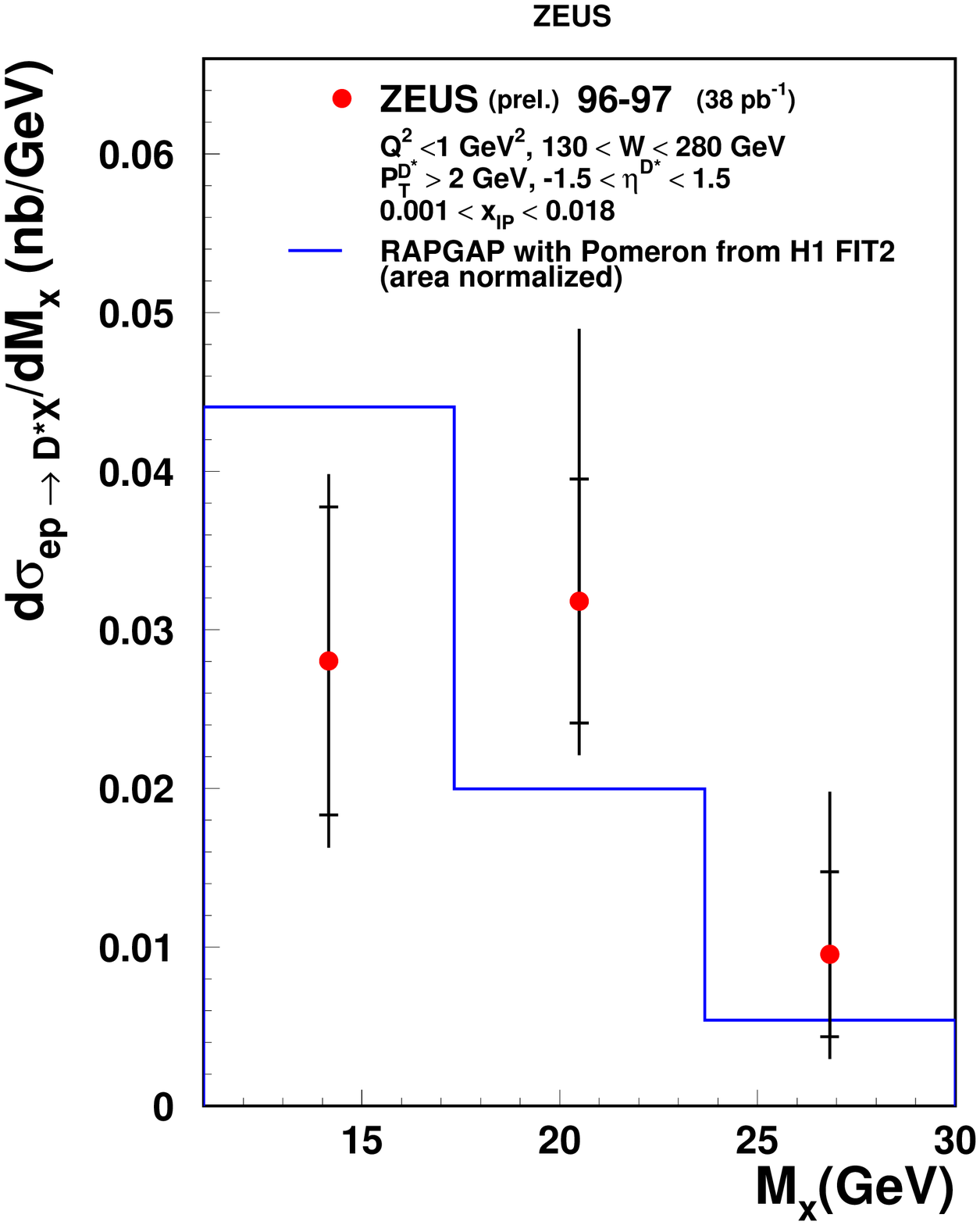}}
                      }% 
\llap{\parbox[t]{27mm}{\vglue53mm\sffamily(c)}}
                                                \nolinebreak
\parbox[t]{0.5\hsize}{\vglue0pt\resizebox*{\hsize}{!}
  {\includegraphics{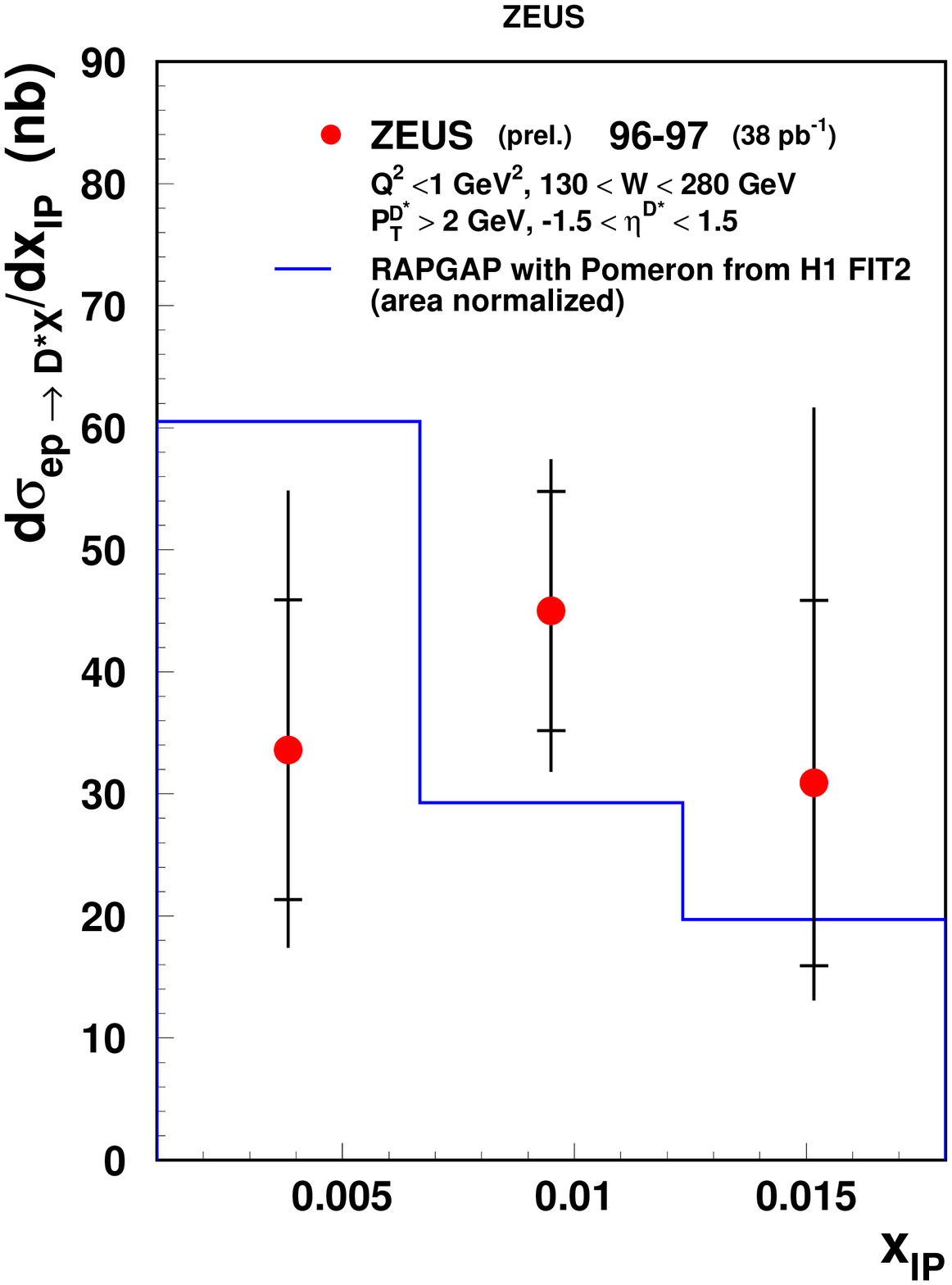}}
                      }%
\llap{\parbox[t]{27mm}{\vglue53mm\sffamily(d)}}   %\par 
\caption{\label{fig:difsigphp}  
Differential cross
 sections (solid points) for the diffractive photoproduction
reaction \mbox{$ep\to D^*Xp$}: 
(a)~${d\sigma}/{dp_T(D^*)}$, 
(b)~${d\sigma }/{d\eta(D^*)}$, 
  (c)~${d\sigma }/{dM_x}$ and (d)~${d\sigma }/{dx_{\pomsub}}$. 
The inner bars show statistical errors, and the outer bars correspond to 
statistical and systematic uncertainties added in quadrature. 
The data are compared with the 
distributions (histogram) of RAPGAP, normalized to the data, 
%MC diffractive events, simulated 
%calculated in the framework of the resolved Pomeron model using the H1 FIT2 Pomeron 
calculated in the framework of the resolved-Pomeron model using the H1 FIT2 Pomeron 
parameterization, which was 
obtained from fits to HERA data.
         }
\end{figure}
\newpage 
\begin{figure}%[t]
\vskip -2.0cm
\parbox[t]{0.5\hsize}{\vglue0pt\resizebox*{\hsize}{!}
  {\includegraphics{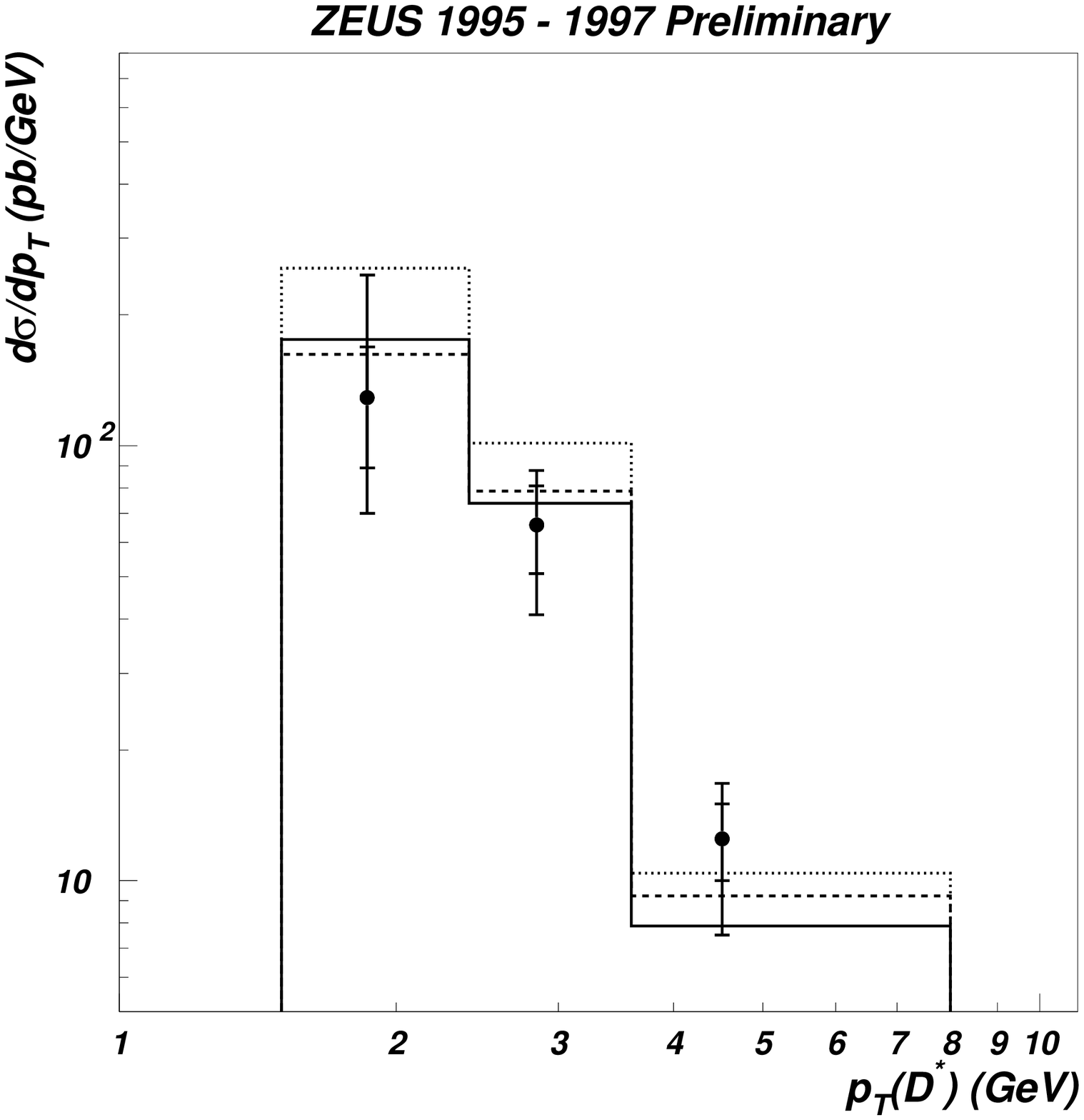}}
                     }%
\llap{\parbox[t]{15mm}{\vglue13mm\sffamily(a)}}
                                               \nolinebreak
\parbox[t]{0.5\hsize}{\vglue0pt\resizebox*{\hsize}{!}
  {\includegraphics{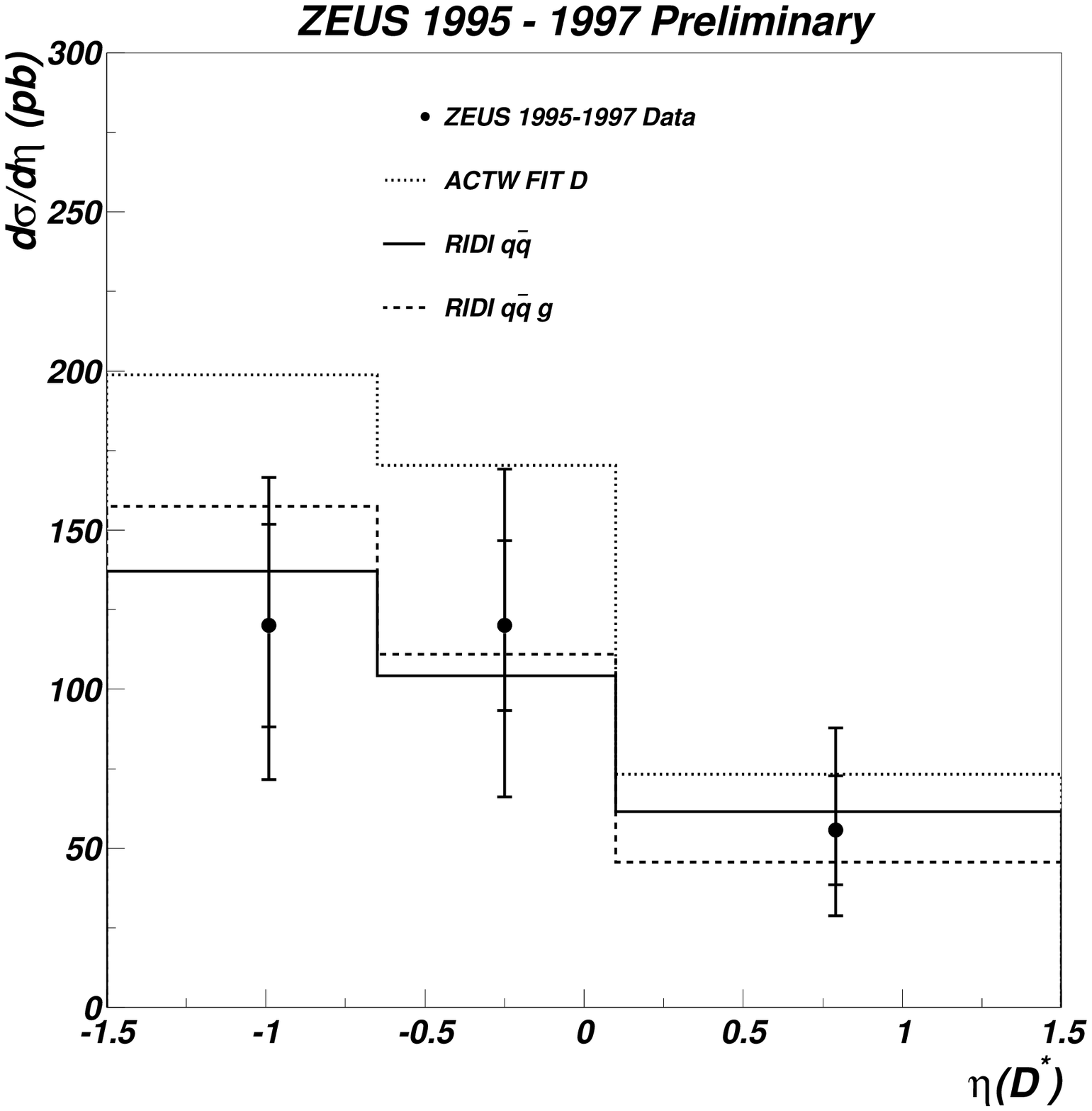}}
                     }%
\llap{\parbox[t]{15mm}{\vglue13mm\sffamily(b)}}\par
\parbox[t]{0.5\hsize}{\vglue0pt\resizebox*{\hsize}{!}
  {\includegraphics{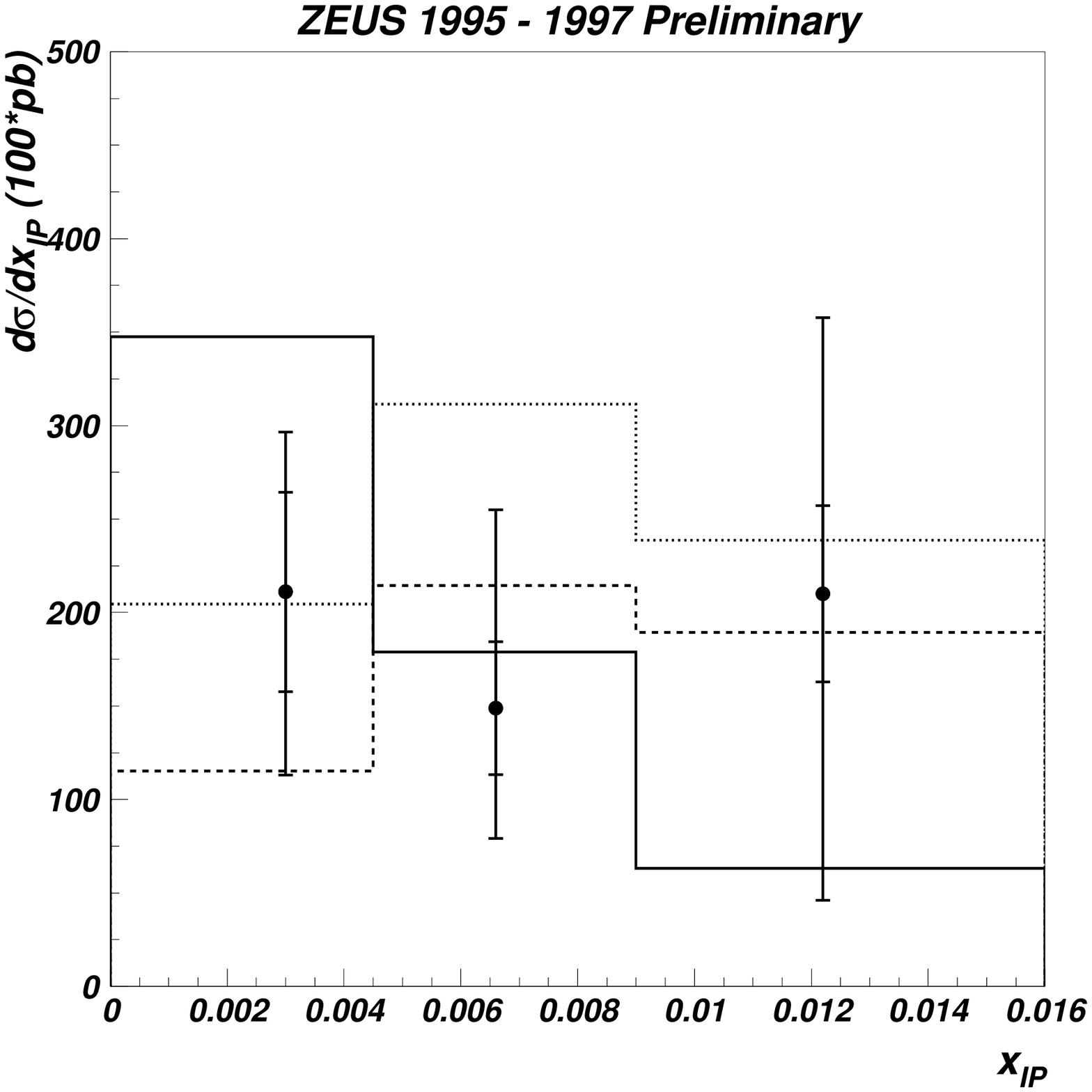}}
                      }% 
\llap{\parbox[t]{15mm}{\vglue13mm\sffamily(c)}}
                                                \nolinebreak
\parbox[t]{0.5\hsize}{\vglue0pt\resizebox*{\hsize}{!}
  {\includegraphics{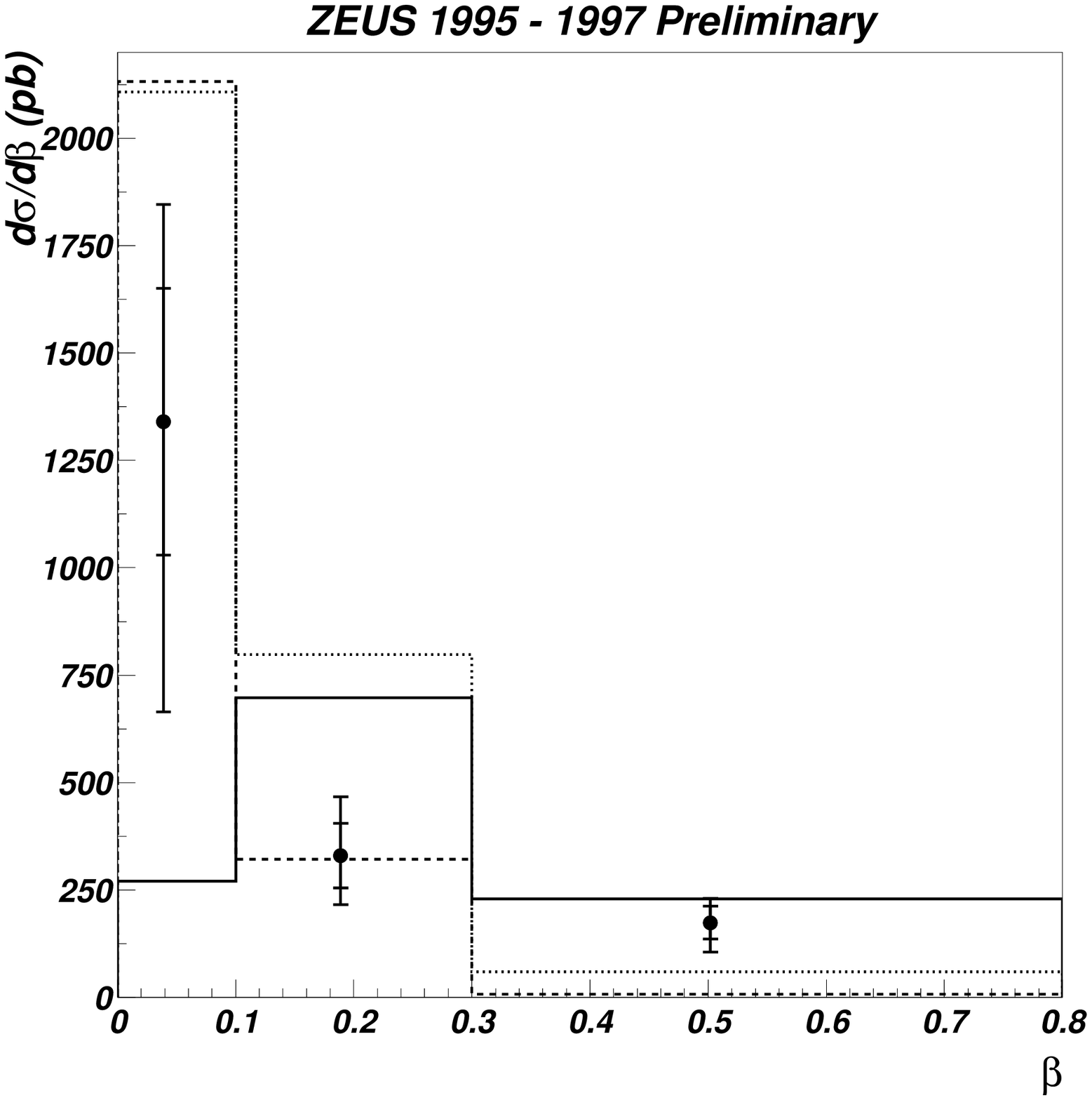}}
                      }%
\llap{\parbox[t]{15mm}{\vglue13mm\sffamily(d)}}   %\par 
\caption{
Differential cross sections (solid points) for the diffractive DIS
reaction \mbox{$ep\to D^{*\pm}Xp$}: 
(a)~${d\sigma}/{dp_T(D^{*\pm})}$, (b)~${d\sigma }/{d\eta(D^{*\pm})}$, 
  (c)~${d\sigma }/{dx_{\pomsub}}$ and (d)~${d\sigma }/{d\beta}$.
The inner bars show statistical errors, and the outer bars correspond to 
statistical and systematic uncertainties 
added in quadrature. The data are compared 
to the  distributions of RAPGAP based on
%MC diffractive events, simulated in the frame of
% resolved Pomeron model with the ACTW  %Pomeron 
the  resolved-Pomeron model with the ACTW  %Pomeron 
parameterization
%, fitted to HERA data 
(dotted histogram). The data are also compared to the distributions of
 RIDI 
% MC diffractive events, simulated in the frame of 
using the two-gluon model
 (solid and dashed histograms representing the $q\bar{q}$ and
              $q\bar{q}g$ components, respectively).  
\label{fig:difsigdis}}
\end{figure}

\begin{thebibliography}{99}
%
\bibitem{D*DIS00} ZEUS Collab., paper 434, subm. to the XXXth Int. Conf. 
on HEP, Osaka, Japan, July 2000. 
%
\bibitem{D*PhP00}I.A. Korzhavina for the ZEUS Collab., materials of the
Rus. Academy of Sci. Nucl. Phys. 2000 Conf., Moscow, Russia, Nov. 2000.
%
\bibitem{p.diss} ZEUS Collab., J. Breitweg  {\it et al.},
                                \Journal{\EPJC}{6}{43}{1999};
                                \Journal{\EPJC}{1}{81}{1998}.                   
%
\bibitem{incD*PhP}ZEUS Collab., J.Breitweg {\it et al.},
                              \Journal{\EPJC}{6}{67}{1999}.
%
\bibitem{resIP}
G. Ingelman and P. Schlein, \Journal{\PLB}{152}{256}{1985}.
%A. Donnachie and P.V. Landshoff, \Journal{\PLB}{191}{}{1987}. 
%A. Donnachie and P.V. Landshoff, \Journal{\NPB}{303}{634}{1988}. 

\bibitem{H1FIT2LO} H1 Collab., C. Adloff {\it et al.},
                                  \Journal{\ZPC}{76}{613}{1997}.

\bibitem{ACTW}L. Alvero {\it et al.}, 
                                 \Journal{\PRD}{59}{74022}{1999};
                                 hep-ph/9806340.
%
\bibitem{RAPGAP}H.Jung, {\em Comp. Phys. Comm.} 86, 147 (1995).
%
\bibitem{2gRIDI} M.G. Ryskin, {\em Sov. J. Nucl. Phys.} 52, 529 (1990).
%
\end{thebibliography}
\end{document}